\documentclass[sigconf]{acmart}
\usepackage{url}
\usepackage{tikz}

\newcommand{\edit}[1] {{\color{black}}{\color{black}{#1}}}

\newcommand{\rebuttal}[1] {{\color{black}}{\color{black}{#1}}}

\usepackage{todonotes}
\usepackage{listings}
\usepackage{xcolor}
\usepackage{algorithm}
\usepackage{algpseudocode}
\usepackage{comment}
\usepackage{hyperref}

\usepackage{mathtools}  
\usepackage{enumitem}
\usepackage{amsthm}
\usepackage{graphicx}

\definecolor{codegreen}{rgb}{0,0.6,0}
\definecolor{codegray}{rgb}{0.5,0.5,0.5}
\definecolor{codepurple}{rgb}{0.58,0,0.82}
\definecolor{backcolour}{rgb}{0.95,0.95,0.92}

\lstdefinestyle{mystyle}{
    backgroundcolor=\color{backcolour},   
    commentstyle=\color{codegreen},
    keywordstyle=\color{magenta},
    numberstyle=\tiny\color{codegray},
    stringstyle=\color{codepurple},
    basicstyle=\ttfamily\footnotesize,
    breakatwhitespace=false,         
    breaklines=true,                 
    captionpos=b,                    
    keepspaces=true,                 
    numbers=left,                    
    numbersep=5pt,                  
    showspaces=false,                
    showstringspaces=false,
    showtabs=false,                  
    tabsize=2
}

\acmConference[SC'25]{The International Conference for High Performance Computing, Networking, Storage, and Analysis}{November 16–21}{St. louis, MO}
\startPage{1}

\newcommand{\tool}[0]{\mbox{\textsc{DynMo}}}

\DeclareRobustCommand*\circled[1]{\tikz[black,baseline=(char.base)]{
            \node[shape=circle,fill,inner sep=1.3pt] (char) {\textcolor{white}{#1}};}}

\title{Balanced and Elastic End-to-end Training of Dynamic LLMs}

\author{Mohamed Wahib}
\affiliation{
  \institution{RIKEN Center for Computational Science}
  \city{Kobe}
  \country{Japan}
}
\email{moahmed.attia@riken.jp}

\author{Muhammed Abdullah Soyturk}
\affiliation{
  \institution{Koç University}
  \city{Istanbul}
  \country{Turkey}
}
\email{muhammetabdullahsoyturk@gmail.com}

\author{Didem Unat}
\affiliation{
  \institution{Koç University}
  \city{Istanbul}
  \country{Turkey}
}
\email{dunat@ku.edu.tr}

\begin{document}

\begin{abstract}

To reduce the computational and memory overhead of Large Language Models, various approaches have been proposed. These include a) Mixture of Experts (MoEs), where token routing affects compute balance; b) gradual pruning of model parameters; c) dynamically freezing layers; d) dynamic sparse attention mechanisms; e) early exit of tokens as they pass through model layers; and f) Mixture of Depths (MoDs), where tokens bypass certain blocks. While these approaches are effective in reducing overall computation, they often introduce significant workload imbalance across workers. In many cases, this imbalance is severe enough to render the techniques impractical for large-scale distributed training, limiting their applicability to toy models due to poor efficiency.

We propose an autonomous dynamic load balancing solution, \tool{}, which provably achieves maximum reduction in workload imbalance and adaptively equalizes compute loads across workers in pipeline-parallel training. In addition, \tool{} dynamically consolidates computation onto fewer workers without sacrificing training throughput, allowing idle workers to be released back to the job manager. \tool{} supports both single-node multi-GPU systems and multi-node GPU clusters, and can be used in practical deployment. Compared to static distributed training solutions such as Megatron-LM and DeepSpeed, \tool{} accelerates the end-to-end training of dynamic GPT models by up to 1.23x for MoEs, 3.18x for parameter pruning, 2.23x for layer freezing, 4.02x for sparse attention, 4.52x for early exit, and 1.17x for MoDs.


\end{abstract}

\maketitle

\section{Introduction}
\label{sec:introduction}

Neural network sizes used to train LLMs have grown exponentially since the introduction of Transformers~\cite{vaswani2017attention}, demanding increasingly more memory and compute power. However, the memory and compute capacity of a single accelerator have not kept pace~\cite{sevilla2022compute}. As a result,  combinations of model and data parallelism have been adopted to train large models~\cite{narayanan2021efficient}.
Pipeline parallelism is one of the most widely used forms of model parallelism in LLMs, where consecutive layers are grouped into stages, each assigned to an accelerator (worker)~\cite{DBLP:conf/hpdc/KahiraNBTBW21}. Input mini-batches are split into micro-batches to increase utilization through pipelined execution~\cite{huang2019gpipe, harlap2018pipedream, fan2021dapple, li2021chimera, qi2024zero}. Pipeline parallelism is essential in practice, as combining data, tensor/expert, and context parallelism alone is insufficient to scale training to tens of thousands of GPUs. Major production families of models such as OpenAI's GPT, Claude, Mixtral, DeepSeek, Qwen, Gemini etc reportedly rely on pipeline parallelism in combination with other forms of parallelism. 

\begin{figure*}[t]
\centering
\includegraphics[width=\textwidth]
{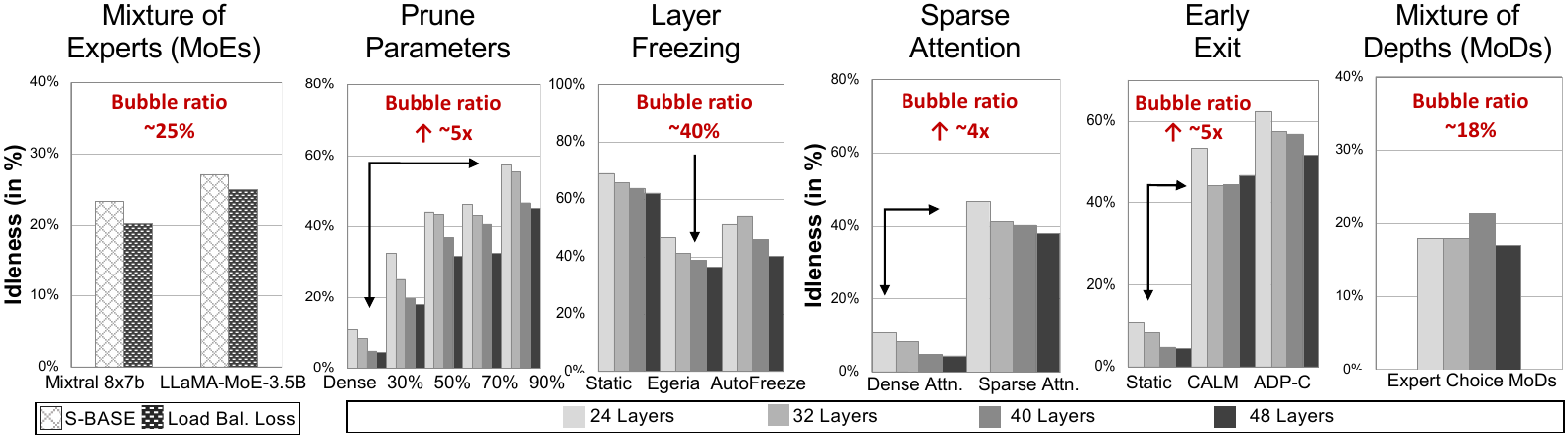}
\caption{
Average idleness percentage of GPUs (per iteration) for training dynamic GPT models~\cite{gpt-2} on 720 H100 GPUs in total, excluding MoEs which uses 128 H100 GPUs in total. We use models parameterized to have between 24 and 48 layers, except for the MoEs user case for which we report the average idleness percentage for Mixtral 8x7b and LlaMA-MoE-3.5B models. For pipeline parallelism, we use the highest performing (single) pipeline parallelism scheme known to the authors: the "almost zero-bubble pipeline parallelism" scheme~\cite{qi2024zero}. The bubble ratios are measured on a hybrid of pipeline and data parallelism. 
\circled{1} \textbf{MoE}: we observe $\sim$25\% bubble ratio in the pipeline on Mixtral 8x7b~\cite{jiang2024mixtral} and LLaMA-MoE-3.5b~\cite{llama-moe-2023}, arising from the load imbalance imposed by the routing schemes used in token choice (S-BASE~\cite{pmlr-v139-lewis21a} and load imbalance with auxiliary loss~\cite{jiang2024mixtral}).
\circled{2} \textbf{Gradual pruning of model parameters}: we observe almost a five fold increase in idleness at 90\% sparsity levels. Note that idleness at \emph{Dense} is the inherent pipeline bubbles of a static model. 
\circled{3} \textbf{Layer freezing}: SoTA freezing schemes that incorporate load balancing (Egeria~\cite{wang2022efficient} and AutoFreeze~\cite{liu2021autofreeze}) yield $\sim$40\% bubble ratio. 
\circled{4} \textbf{Dynamic Sparse Flash Attention}: locality sensitive hashing with support for flash attention~\cite{pagliardini2023fast} exhibits a 4x increase in the bubble ratio over the baseline dense attention.
\circled{5} \textbf{Early exit}: SoTA early exit methods (CALM~\cite{schuster2022confident} and ADP-C~\cite{DBLP:conf/iclr/0003XWDS22}) exhibits up to 5x increase in the bubble ratio over the baseline (w/o early exit), mainly due to the accumulation of bubbles in late layers. 
\circled{6} \textbf{MoD}: we observe $\sim$18\% bubble ratio in the pipeline, arising from the load imbalance imposed by the routing scheme of expert choices that lacks information about future tokens~\cite{Raposo2024MixtureofDepthsDA}.
}

\label{fig:bubble_ratio}
\end{figure*}

In traditional LLM training, each pipeline stage has a fixed workload known in advance. In contrast, emerging schemes designed to reduce computational demands introduce dynamic workloads. These include: neural networks where different input samples take different pathways through the model layers, e.g. gated neural networks~\cite{shazeer2017outrageously}, sparsely activated Mixture of Experts (MoEs)~\cite{DBLP:conf/nips/ZhouLLDHZDCLL22}, Switch Transformers~\cite{fedus2021switch}, Mixture of Depths (MoDs)~\cite{Raposo2024MixtureofDepthsDA} etc.,
 gradual pruning where the parameters of a model are pruned (i.e. sparsified) during training \cite{gale2019state}, 
 freeze training where some of the layers of the model are adapatively frozen during training~\cite{wang2022efficient},  different schemes to dynamically sparsify the attention matrix~\cite{9896137,pagliardini2023fast,tay2020sparse}, and  early exit strategies  where tokens skip remaining layers based on an exit decision~\cite{Elbayad2020Depth-Adaptive,schuster2022confident,DBLP:conf/iclr/0003XWDS22,Kim_2022}. 
 Beyond computational efficiency, dynamic models are also explored for benefits like improved generalization and explainability. We refer readers to~\cite{han2021dynamic,10.1145/3530811} for surveys on dynamic LLMs.

One of the main downsides of using dynamic models is the load imbalance they introduce in pipeline parallelism, which significantly reduces LLM training throughput~\cite{zhou2022mixture, 10.1145/3503221.3508418}. For example, Figure~\ref{fig:bubble_ratio} shows the average GPU idleness for GPT models with varying depths and different dynamic model types. This imbalance appears as {\em bubbles} in the pipeline, where an accelerator stalls while waiting for work to be passed from its neighboring worker. As illustrated in the figure, idleness can range from 18\% to 5x, depending on the scheme, e.g., layer freezing introduces up to 40\% idleness for a 40-layer model.
Note that these bubbles are distinct from the inherent gaps in pipeline scheduling, such as idle periods in the pipeline wind-up and wind-down. The former, which stems from dynamic workloads, is the primary focus of this paper.

Production distributed training frameworks typically apply static load balancing at the start of training and maintain the same distribution throughout. For example, Megatron-LM~\cite{shoeybi2019megatron} evenly splits transformer layers across accelerators. DeepSpeed~\cite{smith_2023} offers three partitioning strategies: \emph{uniform} (equal number of layers), \emph{param} (equal number of parameters), and \emph{regex} (grouping layers by name patterns). These methods assume that workloads remain stable during training. Consequently, they fail to handle the pipeline stalls introduced by dynamic models, leading to reduced computational efficiency.

Innovative designs of dynamic models aim to reduce computational cost, but without effective load balancing, their benefits practically fail to translate into actual performance gains during distributed training~\cite{10643325}. To address this gap, we introduce \tool{}, an elastic load-balancing framework tailored for dynamic models. It is the first work to study pipeline stalls caused by training dynamic models. \tool{} ensures balanced pipeline stages by dynamically redistributing workloads across accelerators whenever imbalance arises, thereby improving computational efficiency and reducing training costs. 
The framework incorporates two different dynamic balancers, both proven to converge to the optimal workload balance among workers. Our experiments demonstrate that \tool{} incurs negligible overhead and scales effectively in both single-node multi-GPU and multi-node multi-GPU setups.

\tool{} also provides the capability to elastically adapt GPU resources. As the total workload decreases during training due to techniques such as gradual pruning or early exit, the load balancer consolidates the work onto fewer GPUs, subject to memory capacity constraints, while maintaining performance. GPUs that are no longer needed can then be released back to the job scheduler. Given the high cost and long duration of LLM training, this elasticity enables significant additional cost savings. Our contributions are:

\begin{itemize}[leftmargin=2.5mm]
  \item 
  We introduce \tool{}, a framework for load balancing dynamic LLMs to improves their end-to-end training efficiency,  making them practical for real-world use. We invoke \tool{} to rebalance at regular intervals without prior knowledge of dynamism, hence balancing the load in a fully automated and transparent manner. 
  \tool{} is orthogonal to the underlying pipeline and dynamism schemes, ensuring compatibility with diverse dynamic compute and model reduction methods. 
  
  \item We propose two load balancing algorithms proven to converge to optimal balancing. 
  Additionally, we introduce a GPU re-packing scheme that reduces the number of GPUs used during training by consolidating work onto fewer devices.
    
  \item 
  We demonstrate the benefits of \tool{} across six dynamic model scenarios in both single-node and multi-node settings. On multi-node hybrid data and pipeline parallelism with up to 720 H100 GPUs, \tool{} delivers speedups of 1.23× (MoEs), 3.18× (parameter pruning), 2.23× (layer freezing), 4.02× (sparse attention), 4.52× (early exit), and 1.17× (MoDs) over static Megatron-LM. Additionally, our re-packing strategy can reduce GPU usage by up to 50\% while maintaining comparable performance.
\end{itemize}

Finally, we emphasize that \tool{} has no impact on model accuracy, as it solely redistributes workload without altering the learning process or regime. \tool{} functions as a complementary system software layer, operating independently of the underlying dynamic strategies such as parameter pruning, early exit, layer freezing, and expert routing. This design makes \tool{} easily extendable and compatible with a wide range of dynamic schemes. In principle, it can also be applied to models that adapt for other reasons, such as hardware variability~\cite{sinha2022not}.
 



\section{Dynamic Models}
\label{sec:background}

Training schemes reducing work can introduce dynamic training workloads. The irregular control-flow in those dynamics models lead inevitably to load imbalance. This leads to inefficiencies that cause the dynamic model to be slower than the baseline, hence defeating the purpose of using a dynamic model. 

The load balancing problem considered can be formally defined as follows. Given a set of workers \( \mathcal{N} = \{\mathcal{N}_1, \mathcal{N}_2, \ldots, \mathcal{N}_n\} \) and a set of tasks \( \mathcal{T} = \{t_1, t_2, \ldots, t_m\} \), each task \( t_j \in \mathcal{T} \) is associated with a workload \( c_j \). The total workload is denoted by:

\small
\[
C = \sum_{j=1}^{m} c_j
\]
\normalsize

Let \( A : \mathcal{T} \to \mathcal{N} \) be a load assignment function that maps each task to a worker. The load of a worker \( \mathcal{N}_i \in \mathcal{N} \), denoted \( L_i \), is defined as the sum of the workloads of tasks assigned to it:

\small
\[
L_i = \sum_{t_j \in A^{-1}(\mathcal{N}_i)} c_j
\]
\normalsize

The objective of the load balancing problem is to minimize the maximum load among all workers:

\small
\[
\min_{A} \max_{i \in \{1, \ldots, n\}} L_i = \min_{A} \max_{i \in \{1, \ldots, n\}} \left( \sum_{t_j \in A^{-1}(\mathcal{N}_i)} c_j \right)
\]
\normalsize

This optimization problem aims to distribute the tasks such that the workload is balanced across the workers, minimizing the worst-case scenario in terms of load. 


In the remainder of this section, we build on the formal load balancing definition by extending the model to capture the dynamism introduced by each of our six example cases. 
In each of the six cases, we define the load on each worker $N_j$ to calculate the load imbalance $\Delta L^{(k)}$ at time step $k$. Ideally, \( L_j^{(k)} \approx L_{j'}^{(k)} \) for all workers, but the dynamic nature of the six cases leads to an imbalance. Defining maximum and minimum loads:

\small
\begin{equation} 
L_{\max}^{(k)} = \max_{j} L_j^{(k)}, \quad L_{\min}^{(k)} = \min_{j} L_j^{(k)}
\label{eq:maxmin}
\end{equation} 
\normalsize

the imbalance is:

\small
\begin{equation}
\Delta L^{(k)} = \frac{L_{\max}^{(k)} - L_{\min}^{(k)}}{\frac{1}{n} \sum_{j=1}^{n} L_j^{(k)}}
\label{eq:imbalance}
\end{equation}
\normalsize

\subsection{Mixture of Experts}

In MoEs~\cite{conf/iclr/ShazeerMMDLHD17}, input tokens are routed to specialized sub-networks (experts) instead of a single feed-forward network, improving efficiency but introducing load imbalance when those experts are distributed among workers. Empirical results from Mixtral 8x7B~\cite{jiang2024mixtral} show up to 25\% imbalance. Prior approaches attempt mitigation via auxiliary losses~\cite{jiang2024mixtral, pmlr-v139-lewis21a}, yet routing remains suboptimal. Recent strategies, such as DeepSeek V3~\cite{deepseekai2025deepseekv3technicalreport}, introduce bias terms in token-to-expert affinity scores~\cite{wang2024auxiliarylossfreeloadbalancingstrategy}, achieving up to ~8\% imbalance per layer in a 3B model. However, imbalance compounds across layers and is expected to be higher in larger models. Load imbalance propagates through training, creating pipeline stalls, especially during \texttt{all\_to\_all} communication. Addressing this remains crucial for efficient MoE training in distributed systems.

Let \( \mathcal{E} = \{e_1, e_2, \dots, e_k\} \) be the set of experts, each assigned to a worker \( \mathcal{N}_j \in \mathcal{N}\). A routing function \( R : \mathcal{T} \to \mathcal{N} \) maps tokens to workers. The load of a worker $L_j$ aggregating the workload $c_{e_j}$ all tokens it is assigned, at time step $k$:

\small
\[
L_j^{(k)} = \sum_{e_j \in \mathcal{E}_i} \sum_{t_k \in R^{-1}(e_j)} c_{ej}^{(k)}
\]
\normalsize

The load per worker defined above can be used with Equations~\ref{eq:maxmin} and~\ref{eq:imbalance} to get the overall system imbalance introduced by the stochastic, or learnable, routing.

\subsection{Parameter Pruning}
Global parameter pruning~\cite{713929} removes a subset of model parameters, creating a sparse network that maintains performance but introduces imbalance due to non-uniform pruning across layers.

Let \( \mathcal{L} = \{l_1, l_2, \dots, l_d\} \) be the set of layers, each assigned to a worker \( \mathcal{N}_j \). During distributed training, workload varies as parameters are pruned. The fraction of retained parameters in layer \( l_i \) at time step \( k \) is \( p_i^{(k)} \), the total network compute being $C^{(k)}$, where \( A^{(k)} : \mathcal{L} \to \mathcal{N} \) be the layer-to-worker assignment. The load on worker \( \mathcal{N}_j \) is:

\small
\[
L_j^{(k)} = \sum_{l_i \in (A^{(k)})^{-1}(\mathcal{N}_j)} \sum_{i=1}^{d} p_i^{(k)} \cdot c_i
\]
\normalsize

The load per worker defined above can be used with Equations~\ref{eq:maxmin} and~\ref{eq:imbalance} to get the overall system imbalance introduced by the pipeline stalls occurring while some workers wait for heavily pruned layers to complete~\cite{zhu2017prune, frankle2018lottery, bellec2017deep}. 

\subsection{Layer Freezing}

Layer freezing reduces computational costs by halting updates to certain layers once they have converged. While improving efficiency, it introduces load imbalance when frozen layers are unevenly distributed among workers~\cite{shen2020reservoir}.

Let \( \mathcal{L} = \{l_1, l_2, \dots, l_d\} \) be the set of layers in an LLM, each assigned to a worker \( \mathcal{N}_i \). Earlier layers often converge faster~\cite{wang2022efficient}, enabling their freezing to save computation. Let \( f_i^{(k)} \in \{0,1\} \) indicate whether layer \( l_i \) is frozen at time step \( k \):

\small
\[
f_i^{(k)} = 
\begin{cases}
1, & \text{if layer } l_i \text{ is frozen at } k, \\
0, & \text{otherwise}
\end{cases}
\]
\normalsize

For \( c_i \) being the initial compute. If \( f_i^{(k)} = 1 \), then \( c_i^{(k)} = 0 \), contributing no computational load. For the total workload of layer \( l_i \) at time step \( k \) is, where \( A^{(k)} : \mathcal{L} \to \mathcal{N} \) is the layer-to-worker assignment. The load on worker \( \mathcal{N}_j \) is:

\small
\[
L_j^{(k)} = \sum_{l_i \in (A^{(k)})^{-1}(\mathcal{N}_j)} \sum_{i=1}^{d} (1 - f_i^{(k)}) \cdot c_i
\]
\normalsize

The load per worker can be used with Equations~\ref{eq:maxmin} and~\ref{eq:imbalance} to get the overall system imbalance introduced by uneven freezing.



\subsection{Dynamic Sparse Flash Attention} 
Dynamic sparse flash attention combines hash-based sparse attention with FlashAttention~\cite{10.5555/3600270.3601459, pagliardini2023fast}, accelerating computations by restricting the attention matrix to blocks determined by hash codes. However, varying sparsification across layers causes load imbalances during distributed training.

Let \( \mathcal{L} = \{l_1, l_2, \dots, l_d\} \) be the set of layers, each using dynamic sparse attention. The workload depends on the sparsity level, which changes dynamically. Let \( s_i^{(k)} \in [0,1] \) be the sparsity factor of layer \( l_i \) at time \( k \), representing the fraction of nonzero elements in the attention matrix. For \( c_i \) being the initial workload of layer \( l_i \). Let \( A^{(k)} : \mathcal{L} \to \mathcal{N} \) be the layer-to-worker assignment. The load on worker \( \mathcal{N}_j \) is: at time \( k \) is:

\small
\[
L_j^{(k)} = \sum_{l_i \in (A^{(k)})^{-1}(\mathcal{N}_j)} \sum_{i=1}^{d} s_i^{(k)} \cdot c_i
\]
\normalsize

The load per worker defined above can be used with Equations~\ref{eq:maxmin} and~\ref{eq:imbalance} to get the overall system imbalance introduced by sparsity, which varies across layers and time steps.




\subsection{Early Exit}
Early Exit allows tokens to skip layers once they reach a confident state, reducing computation but causing load imbalance due to uneven token distribution across layers. This can increase bubble ratios by up to 5x due to idle time~\cite{schuster2022confident}.

Let \( \mathcal{L} = \{l_1, l_2, \dots, l_d\} \) be the set of layers where tokens can exit early. Let \( t^{(k)} \) be the total tokens at time \( k \), and \( t_i^{(k)} \) as the tokens processed by layer \( l_i \). Typically, \( t_i^{(k)} \) decreases with depth~\cite{schuster2022confident,DBLP:conf/iclr/0003XWDS22}. For \( c_i \) being the initial workload. For $c_i$ being the initial workload, the total workload of all layers is:

\small
\[
C^{(k)} = \sum_{i=1}^{d} \frac{t_i^{(k)}}{t^{(k)}} \cdot c_i
\]
\normalsize

Let \( A^{(k)} : \mathcal{L} \to \mathcal{N} \) be the layer-to-worker assignment. The load of worker \( \mathcal{N}_j \) is:

\small
\[
L_j^{(k)} = \sum_{l_i \in (A^{(k)})^{-1}(\mathcal{N}_j)} c_i^{(k)}
\]
\normalsize

The load per worker defined above can be used with Equations~\ref{eq:maxmin} and~\ref{eq:imbalance} to get the overall system imbalance introduced by reduced work in later layers of the model.


\subsection{Mixture of Depths} 
MoD~\cite{Raposo2024MixtureofDepthsDA} generalizes early exit by allowing tokens to skip both intermediate and final layers. The MoD variant used in this work employs expert choice via MoEs for improved performance. However, dynamic layer skipping and variability in expert selection introduce significant load imbalances in distributed training. Empirical results show up to 18\% imbalance in MoDs ~\cite{Raposo2024MixtureofDepthsDA}.

Let \( \mathcal{L} = \{l_1, l_2, \dots, l_d\} \) be the set of layers, with \( t^{(k)} \) as the total number of tokens at time \( k \), and \( t_i^{(k)} \) as the tokens processed by layer \( l_i \). MoD enables tokens to skip layers based on expert predictions, leading to fluctuating workloads. The routing weight \( r_i^{(k)} \) determines whether tokens bypass a layer, and the total effective compute of \( l_i \) at time \( k \) is:

\small
\[
C^{(k)} = r_i^{(k)} \cdot t_i^{(k)} \cdot c_i
\]
\normalsize

Let \( A^{(k)} : \mathcal{L} \to \mathcal{N} \) be the load assignment function, then each worker \( \mathcal{N}_j \) would have the load:

\small
\[
L_j^{(k)} = \sum_{l_i \in (A^{(k)})^{-1}(\mathcal{N}_j)} c_i^{(k)}.
\]
\normalsize

The load per worker defined above can be used with Equations~\ref{eq:maxmin} and~\ref{eq:imbalance} to get the overall system imbalance introduced by: a) the MLP predictor's misestimation of whether a token will be among the top-k selected for the next layer, and b) possible imbalance from the underlying MoE that the MoD sits on top of.



        

        
\section{\tool{} Framework}
\label{sec:design}


\begin{figure}[t]
  \includegraphics[width=\linewidth]{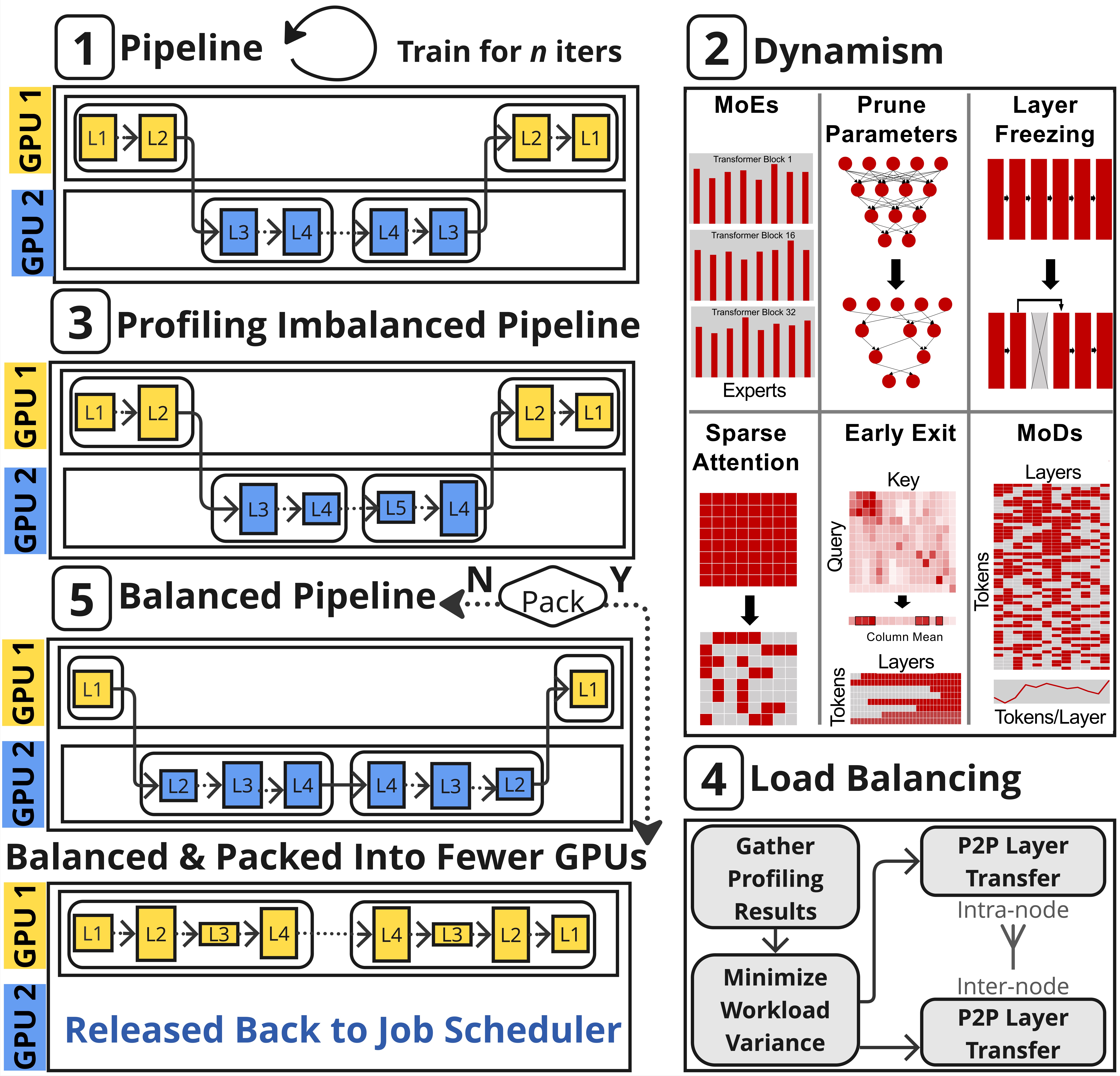}
  \caption{Overview of \tool. The flow in the figure (top to bottom) is repeated until training is completed. Each yellow and blue rectangle represents a transformer layer. The size of a rectangle illustrates the amount of work in a layer. (1) shows the pipeline before the model starts to change due to dynamism. (2) some action (dynamically)  changes the model or the flow of work inside the model changes. (3) profiles the pipeline to check if there is any imbalance between stages, (4) performs load balancing based on the profiling results, (5) trains the balanced pipeline until the next time to rebalance, optionally it reduces the number of resources (GPUs) used in training by re-packing. 
  }
  \label{fig:flow}
\end{figure}
\subsection{Overview}
We evaluate our load balancing system on six representative dynamic model cases that static distributed training systems struggle to handle efficiently. While our work focus on these cases, the approach generalizes to other forms of dynamic models.


Figure \ref{fig:flow} illustrates the overview of \tool{} with all its steps. The implementation of individual steps of model (or control-flow) altering, load balancing, and re-packing can be found in their respective sections. The dynamism depends on the target case. For instance, if the target case is parameter pruning, the dynamism function would apply global pruning on different worker. 

\tool{} takes as input a model, the number of training iterations, and several dynamism configuring arguments (that have default values if the user wishes to use \tool{} out-of-the-box). 
We start the training with the original model and train it until the dynamism to apply is reached. The frequency varies by the target case, in layer freezing it is as frequent as every 50 iterations, while in parameter tuning the pruning frequency is in the range of 3000-7000 iterations. The model or control flow is altered only if the training is in the dynamism stage. In this stage, the model is adjusted by either being modified or when the control flow inside the model changes, until it meets the training stopping criteria.
The first iteration after each dynamism operation is used for profiling the time it takes to execute each layer in the altered model and the memory usage of all workers (accelerators) in the pipeline. Next, \tool{} collects the profiling information and decides on balancing the workload by moving layers across pipeline stages based on the execution times of individual layers to minimize the pipeline stalls, subject to the constraints of memory capacity per worker. \tool{} also attempts to re-pack the total workload into fewer number of GPUs if the re-packing feature is enabled by the end user. Once the training is out of the dynamism region, the  balanced pipeline continues to execute with the model. 

 By large, the mechanism for measuring the load imbalance, redistributing the load, and re-packing (when possible) does not vary from case to case. However, we need to identify the load imbalance arising from dynamism. For each of the six example cases, a representative case is discussed next.
 

\subsection{Imbalance Caused by Dynamism in Model Training}
\tool{} operates as a black-box approach where the load balancing happens at regular fixed intervals, without any knowledge of whether the model has changed or not. As will be shown in the results section, the very low overhead allows \tool{} to be invoked even at the granularity of each iteration. Due to space constraints, we focus here on describing the  imbalance due to dynamism for gradual pruning, as a representative of the different cases we cover in this paper.

\subsubsection{\textbf{Gradual Global Magnitude Pruning}}
For our pruning design, we use the gradual pruning schedule proposed in~\cite{zhu2017prune} which is formulated as: 

\small
\begin{equation}
 S_t = S_f + (S_i - S_f)(1 - \frac{t - t_0}{n \Delta t})^3, \mathrm{~~~~~~} t \in \{t_0, t_0 + \Delta t, ..., t + n \Delta t\}
\end{equation}
\normalsize

where \(S_i\), \(S_f\), \(n\), \(t_0\), and $\Delta t$ are initial sparsity, final sparsity, number of pruning steps, initial pruning step, and pruning frequency, respectively. The aim of this schedule is to prune the model rapidly in the initial pruning steps since there are many irrelevant connections, then reduce the pruning rate as the number of parameters in the network gets smaller.

\begin{algorithm}[t]
\caption{Dynamism (Global Pruning as an Example)}
\label{alg:global_pruning_algo}
\begin{algorithmic}[1]
    \Statex \textbf{Input:} model, sparsity, rank
    \Statex \textbf{Output:} model
    \State params $\gets$ concat\_params(model)
    \State k $\gets$ num\_params $\times$ (1 - sparsity)
    \State local\_topk, local\_topk\_indices $\gets$ topk(abs(params), k)

    \State topk\_values $\gets$ gather(local\_topk)
    
    \If {rank == 0}
        \State global\_topk\_indices $\gets$ topk(abs(topk\_values), k)
    \EndIf
    \State indices\_to\_keep $\gets$ scatter(global\_topk\_indices)
    \State model = compress\_model(model, indices\_to\_keep)
    \State $\textbf{return}$ model
\end{algorithmic}
\end{algorithm}

We employed an unstructured magnitude pruning technique as opposed to a structured one since unstructured magnitude pruning typically retains better accuracy under high sparsity rates~\cite{prasanna2020bert}. Unstructured magnitude pruning is applied globally (taking all parameters in the model into account) instead of locally since it has been empirically shown that global pruning yields better accuracy under high compression ratios~\cite{blalock2020state}.

To our knowledge, there is no deep learning framework that supports global pruning on a distributed model at the time of this writing (support is only for undistributed models). Thus we implemented our own global pruning algorithm as shown in Algorithm~\ref{alg:global_pruning_algo}. The global pruning method takes three arguments, namely the model, target sparsity, and the rank of the device. Note that each rank\footnote{We use one MPI rank per GPU.} has only its own portion of the model. First, each rank finds its own local top-$k$ parameters in terms of magnitude (line 3). Then, rank 0 gathers the top-$k$ parameters
of each rank (line 4).
When rank 0 receives all top-$k$ parameters, it calculates the indices of global top-$k$ parameters to keep (line 6), and sends the indices that belong to each rank (line 8). Finally, after each rank receives its indices to keep, they prune (discard) parameters with all other indices in their local parameters (line 9).

\subsection{Load Balancing}
\tool{} implements two load balancing algorithms, and can be extended to support other algorithms. The first is centralized parameter-based partitioning that balances partitions based on the number of parameters. The load balancing algorithm is built on top of DeepSpeed's load balancing utility functions for partitioning in model parallelism~\cite{smith_2023}.
The second algorithm is an iterative decentralized diffusion-based algorithm that aims to minimize the variance between the workload of each rank by attempting to move layers from overloaded workers to underloaded ones in an iterative way. The workload can be described by either the layer execution times, or the parameter counts as in the centralized partitioning method. 

We demonstrate that the two load balancing schemes meet the goals for optimal load balancing by using the following lemmas. 


\textbf{Lemma 1.} \textit{A centralized load balancer $L_c$ over $\mathcal{N}$ workers satisfies maximum reduction in the imbalance $\mathcal{N}_i$ if and only if $\mathcal{N}_i$ reduces the bubble ratio to minimum.}

\textit{Proof.} We will prove by contradiction. Suppose a centralized load balancer $L_c$ over $\mathcal{N}$ workers satisfies maximum reduction in the imbalance $\mathcal{N}_i$ when $\mathcal{N}_i$ has a bubble ratio higher then the minimum. By the definition of maximum reduction in load balance, $L_c$ must preserve minimum differential between the loads of workers $\mathcal{N}_i$ and $\mathcal{N}_j$, which $\mathcal{N}_i$ and $\mathcal{N}_j$ have the minimum load and maximum loads in $\mathcal{N}$, respectively. Consequently, increasing the bubble ratio of $\mathcal{N}_j$ changes the load difference between $\mathcal{N}_i$ and $\mathcal{N}_j$. This is in contradictory of $L_c$ achieving maximum reduction on imbalance.
\qedsymbol

\textbf{Lemma 2.} \textit{An iterative decentralized diffusion based load balancer $L_d$ over $\mathcal{N}$ workers satisfies maximum reduction in the imbalance $\mathcal{N}_i$ if and only if $\mathcal{N}_i$  reduces the bubble ratio to minimum. Also the load balancer is guaranteed to converge to the maximum reduction in imbalance in the following number of rounds}

\small
\begin{align*}
O\left ( \mathrm{min}\left \{ \mathcal{N}^2 \mathrm{log}\left ( \frac{S\mathcal{N}}{\gamma} \right ) \mathrm{log}~\mathcal{N}, \frac{S\mathcal{N}~\mathrm{log}~\mathcal{N}}{\gamma}\right  \}  \right )
\end{align*}
\normalsize

where $\gamma \in \mathbb{R}_{> 0}$ is the convergence factor and $\in \mathbb{R}_{> 0}$ is the total number of stages in the pipeline.

\textit{Proof.} We leverages core ideas from Lyapunov optimization. We first define a potential function, $\phi$, that measures at each round the total magnitude of workload gaps in the system:
\begin{align*}
    \forall r\geq 0: \phi (r) = \sum_{u,v\in V}\left | x_u(r)- x_v(r) \right |
\end{align*}
Similar to a Lyapunov function, $\phi$ maps the system state (in this case, a vector of workloads for $\mathcal{N}$ workers) at any given round to a non-negative scalar value that describes the desirability of the current system state. As $\phi$ decreases toward $0$, the system state becomes more desirable; i.e. the workload is balanced across $\mathcal{N}$. As in a standard Lyapunov optimization, we show below that the modifications to a system state caused by executing a single round of our max neighbor algorithm will drift the value of $\phi$ toward zero in a non-decreasing manner. We establish a probabilistic lower bound for the amount of drift in a given round to obtain our time bounds. 

For a given round $r\geq 0$ and node pair $u,v\in V$ , we define $d_{u,v}(r)=\left | x_u(r)- x_v(r) \right |$ to describe the gap
between $u$ and $v$’s workload at the end of that round. For each such $r$, we also define: 
$\left \{ \left \{ u,v \right \} | u \text{~and~} v \text{~connected and averaged their workloads in round~} r\right \}$, i.e., the set of node pairs that connect and average in $r$, and $D_r=\sum_{u,v\in A_r}d_{u,v}(r-1)$, i.e., the sum of gaps averaged in $r$. Finally, we define $t_{max}(r)=max_{u,v\in V}\left \{ d_{u,v}(r) \right \}$ to describe the largest gap between any two nodes at the end of round $r$. From the above analysis that $\phi(r)$ decreases by at least $D_r$ in each round $r$, we proceed to prove the converge time complexity bound.

For a maximum number of rounds to converge to the minimum imbalance:
\begin{align*}
O\left ( \mathrm{min}\left \{ \mathcal{N}^2 \mathrm{log}\left ( \frac{S\mathcal{N}}{\gamma} \right ) \mathrm{log}~\mathcal{N}, \frac{SN~\mathrm{log}~\mathcal{N}}{\gamma}\right  \}  \right )
\end{align*}
Note that these two bounds essentially coincide at $\tilde{O}(\mathcal{N}^2)$ with $\gamma = \Theta (S/n)$, where the notation $\tilde{O}$ hides logarithmic factors. In other words, if we want all nodes to have the same workload up to a constant factor, the max neighbor strategy uses $\tilde{O}(\mathcal{N}^2)$ rounds. We first note that if we arrive at a round r in which $\phi(r)\leq \gamma$, then the system ends this round $\gamma$-converged, i.e. the sum of the gaps is at most $\gamma$, and thus clearly any individual gap is at most $\gamma$. Since $\phi$ is monotonically non-increasing, it follows that every round ${r}'\geq r$ is also $\gamma$-converged. So we just need to show that with high probability, $\phi$ will decrease to $\gamma$ in the time bound stated by the theorem statement.

For each $r\geq 1$, we call r “good” if and only if $\phi(r-1)-\phi(r)\geq s_{max}(r-1)/(60~ln (2n))$. We next calculate how many good rounds guarantee that $\phi$ falls below $\gamma$. To do so, we first note that, non-good rounds cannot increase $\phi$, so we are safe to focus only on reductions generated by good rounds in calculating our bound.

By the definition of $\phi$, for each $r\geq 1$ we know that $\phi(r)< s_{max}(r)n^2$. It follows that if $r$ is a good round, then it decreases $\phi(r-1)$ by a multiplicative factor less than $(1-\frac{1}{60n^2ln(2n)})$. Finally, we also observe that $s_{max}(0)\leq S$ and therefore $\phi(0)< Sn^2$. Leveraging these observations, to find the number of good rounds needed to decrease $\phi$ below $\gamma$, we just need to find the minimum $s$ time steps such that
\begin{align*}
    Sn^2\left (1-\frac{1}{60n^2ln(2n)}\right )\leq \gamma
\end{align*}
A simple calculation implies that $s_{con}=60n^2ln(2n)ln(Sn^2\gamma^{-1})$ is sufficient to satisfy this inequality. We have now established that after $s_{con}$ good rounds the system will be $\gamma$-converged for all future rounds. We are left to bound the number of rounds required to generate $s_{con}$ good rounds with high probability.

For each round $r$, let $X_r$ be the random indicator variable that evaluates to $1$ if round $r$ is good and otherwise evaluates to $0$. We know a given round $r$ is good with probability at least $1/\mathcal{N}$, regardless of the history of the execution through the round $r-1$. We cannot, however, directly leverage this observation to calculate (and concentrate) the expected sum of $X$ variables for a given execution length, as the distribution determining a given $X_r$ might depend in part on the outcome of previous experiments. To overcome this issue, we define for each round $r$, a trivial random indicator variable $\hat{X}_r$ that evaluates to $1$ with independent probability $1/N$ and otherwise evaluates to $0$. Note that for each $r$, $X_r$ stochastically dominates $\hat{X}_r$, and therefore for each $s> 0,Y_s=\sum_{r=1}^{s}X_r$ stochastically dominates $s>0,\hat{Y}_s=\sum_{r=1}^{s}\hat{X}_r$. It follows for any $s> 0$, if $\hat{Y}_s\geq s_{con}$ with some probability $p$ then $Y_s\geq s_{con}$ with probability at least $p$.

A Chernoff bound applied to $\hat{Y}_s$, for $s=c.s_{con}$ (where $c \geq 1$ is a sufficiently large constant defined with respect
to the constants in $s_{con}$ and the constants in the Chernoff form used), provides that $\hat{Y}_s\geq s_{con}$ with high probability,
and therefore so is $Y_s$. To conclude the proof, we note that $c.s_{con}\in O\left ( \mathcal{N}^2log(\frac{S\mathcal{N}}{\gamma})log\mathcal{N} \right )$, as required by the theorem $\gamma$ statement.
\qedsymbol

The diffusion-based load balancing algorithm achieves ideal load balancing by iteratively minimizing workload imbalances using a Lyapunov-inspired approach. The potential function $\phi$, defined as the sum of workload gaps between workers, serves as a measure of imbalance in the system. Each iteration of the algorithm reduces $\phi$ by redistributing tasks from overloaded to underloaded workers, prioritizing layer transfers that yield the largest reductions in imbalance while satisfying memory constraints. The algorithm's probabilistic analysis guarantees that $\phi$ decreases towards a convergence threshold $\gamma$, with the rate of convergence bounded by $O(\mathcal{N}^2 \log(S\mathcal{N}/\gamma) \log\mathcal{N})$, where $\mathcal{N}$ is the number of workers and $S$ is the total pipeline size. This systematic reduction ensures that workload imbalances, quantified by the bubble ratio, are minimized, driving the system towards an optimally balanced state. The theoretical guarantees of convergence and imbalance reduction underpin the algorithm's robustness in dynamic environments.

\subsubsection{\textbf{Overhead of \tool{} and Frequency of Dynamism}}
\tool{} incurs negligible overhead (detailed in the evaluation section). Across all model sizes and dynamism scenarios, total overhead stays within a few single-digit percentages. This includes profiling, the rebalancing algorithm, and inter-GPU layer migration.

As a general rule, we apply rebalancing via \tool{} whenever the model or control flow changes. Given its low overhead, \tool{} can be used as frequently as needed, depending on application requirements. For example, in gradual pruning, dynamism (that is, model changes requiring rebalancing) typically occurs every few thousand iterations. In contrast, for MoEs and MoDs, rebalancing is needed every iteration due to unpredictable imbalances caused by routing decisions taken during the forward pass at each FFN. In these cases, we perform rebalancing during backpropagation, coupling layer migration with the pipeline parallelism scheme by moving layers while the gradients calculation take place, from the last to the first layer.

\subsection{Re-packing the Model to Fewer Workers}

Workload re-packing is the process of consolidating the total training workload onto a reduced number of workers (GPUs) when the overall compute demand drops, allowing idle GPUs to be released. 
For long training schedules common in LLM training, this elasticity can yield significant cost savings. It can also lead to improved performance by decreasing the number of cross-worker communications and mitigating pipeline bubbles.

\begin{algorithm}[t]
\caption{Re-pack Layers into Fewer Workers}
\label{alg:repack_workloads}
\begin{algorithmic}[1]
\Statex $\textbf{Input:}$ active\_workers, mem\_usage 
\Statex $\textbf{Input:}$ target\_num\_workers, num\_layers
\Statex $\textbf{Output:}$ transfers (list)
\State transfers $\gets$ []

\For{src in range(num\_ranks)}
    \For{dst in range(src + 1, num\_ranks)}
        \If{mem\_usage[src] + mem\_usage[dst] $<$ MAX\_MEM  \\
 \quad \space  \quad \space
        \&\& sum(active\_workers) $>$ target\_num\_workers}
            \State active\_workers[src] = 0
            \For{lyr\_idx in range(num\_layers[src])}
                \State transfers.append((src, dst, lyr\_idx))
            \EndFor
            \State mem\_usage[dst] += mem\_usage[src]
            \State num\_layers[dst] += num\_layers[src]
        \EndIf
    \EndFor
\EndFor

\State $\textbf{return}$ transfers
\end{algorithmic}
\end{algorithm}


\subsubsection{\textbf{Re-packing Algorithm}}
As shown in Algorithm \ref{alg:repack_workloads}, we implement workload consolidation using a first-fit algorithm  with the objective of reducing the number of active workers, subject to memory capacity constraints. The algorithm iteratively examines each pair of workers (lines 2-3): if their combined memory usage fits within a single GPU’s memory budget (lines 4-5), the workload is migrated to consolidate onto one GPU, freeing the other (lines 7-8). This process continues across the set of workers until no further consolidation is possible. 
Re-packing is scheduled at the end of each training iteration, leveraging the existing synchronization barrier to avoid the need for introducing additional barriers.




\subsubsection{\textbf{Releasing Unused GPUs after Re-packing}}
To ensure the GPUs are released in a practical manner, we use the NCCL communicator. 
While the NCCL communicator can not be resized, NCCL supports splitting the communicator via \verb!ncclCommSplit()!  into multiple sub-partitions, or even creating a single communicator with fewer ranks~\cite{nccl-com}. After re-packing, active GPUs are assigned to a new sub-communicator, while idle GPUs are assigned to a different one. Since idle GPUs are excluded from the active communicator, concurrent communicators can safely proceed without the risk of deadlock.

Alternatively, re-packing can be coordinated with checkpointing. Unlike rebalancing, which may occur every few iterations, re-packing is infrequent—typically triggered after thousands of iterations, when the model structure has changed significantly. By combining re-packing with a checkpoint restart, the implementation is simplified since a new NCCL communicator is already created during the restart. Moreover, because the model is reloaded and reshared among the workers during checkpoint recovery, there is no additional overhead for resharding the model to a new set of workers.


\tool{}'s supports the release of the GPUs to the ECK (Elastic Cloud on Kubernetes)~\cite{k8}. After re-packing, \tool{}, signals the release of GPUs release back to the ECK-managed Kubernetes cluster. In particular \tool{} uses the Kubernetes API to send a {\tt PATCH} request to the Kubernetes API server to update the {\tt resources.requests} and {\tt resources.limits} fields in the pod spec, effectively reducing or removing GPU resource claims. ECK, which automates the orchestration of containerized workloads and their underlying resources, then detects these freed GPUs and reassigns them to other pending jobs in the queue. This integration allows for more efficient and elastic resource allocation across workloads, minimizing idle GPU time and improving overall utilization.

Lastly, even though \tool{}'s speedup mostly comes from dynamic load balancing, re-packing provides 4–11\% additional performance benefit from the reduced number of number of ranks and volume of communication.
\section{Implementation}
\label{sec:implementation}
The \tool{} framework was developed on top of NVIDIA Megatron Core 0.9.0\footnote{\url{https://github.com/NVIDIA/Megatron-LM/releases/tag/core_r0.9.0}}. Each component of \tool{}, namely hooking to dynamism point in model training, load balancing, and re-packing is implemented in a separate package for ease of use and extension.

The {\tt gather} and {\tt scatter} operations in global pruning were implemented by employing NCCL Peer-to-Peer (P2P) {\tt send-receive} operations instead of collectives since the sizes of the objects to be sent ({\tt local\_topk}) and received ({\tt indices\_to\_keep}) in Algorithm 
\ref{alg:global_pruning_algo} from each rank are different and other ranks do not have this size information to participate in the collective call. It is worth mentioning that while an {\tt alltoallv} collective could be used as an alternative, using it however would add unnecessary global synchronization and may cause contention. Thus we opt for P2P communication.

The necessary information for load balancers such as layer execution times and memory usage comes from the profiling iteration after each pruning iteration. The execution time profiling is implemented by extending the built-in timers of Megatron-LM. The memory consumption of each pipeline stage is gathered with PyTorch's memory statistics for CUDA.

\subsection{Dynamic Reconfiguration of the Pipeline}
DeepSpeed's PipelineModule~\cite{deepspeed_pipe} currently offers three partitioning strategies for distributing model layers, which can be set using the \verb!partition_method()! keyword argument passed to the \verb!PipelineModule!. This allows us to move the layers between GPUs, when needed. When a layer is migrated from GPU $A$ to GPU $B$, the memory allocated for the layer (weights, activation, gradients, optimizers) is released on GPU $A$ and allocated on GPU $B$. As the training resumes (after the layers are migrated), the underlying pipeline scheme starts to pass the new mini-batches along the new distribution of layers in the pipeline.

\subsection{Implementations of Dynamic Models}
\subsubsection{\textbf{Mixture of Experts}}
We build on Mixtral 8x7b weights from Hugging Face~\cite{mixtra} and LLaMA-MoE-3.5B~\cite{llama-moe-2023} in continual training by monitoring the imbalance between layers with respect to number of assigned tokens. We used both the auxiliary load balance scheme adopted in Mixtral 8x7b~\cite{jiang2024mixtral} and the S-BASE load balancer~\cite{lewis2021base}. Since the imbalance is dependent on the point the routing decision is taken, i.e., in the forward pass at each FFN, we rebalance in the back propagation phase where we attach the movement of layers to the pipeline parallelism communication flow.

\subsubsection{\textbf{Parameter Pruning}}
Among the six example cases, gradual pruning posed a unique challenge, as it required code changes to integrate with \tool{}. Unstructured pruning demands a sparse storage format for efficient training and transfer. A common choice is the compressed sparse row (CSR) format, which necessitates replacing dense matrix multiplications (DMM) with sparse equivalents (SpMM). 
Since PyTorch does not support computing the derivative of SpMM operations for backpropagation on a CSR tensor, we evaluated GPU-compatible CSR-based SpMM implementations, namely cuSPARSE by Nvidia and Sputnik~\cite{gale2020sparse}. 
Sputnik's SpMM consistently outperformed cuSPARSE across all tested sparsity levels. This is largely due to Sputnik's kernels being tailored for deep learning workloads, unlike cuSPARSE, which is optimized for HPC workloads with extreme sparsity (often >99\%). Notably, Sputnik begins to outperform cuBLAS around 75\% sparsity. 
Thus, for sparse operations, we implemented PyTorch bindings for the CUDA kernels of Sputnik.

\subsubsection{\textbf{Layer Freezing}}
\tool{} sits on top of layer freezing solutions. More specifically, we build on Wang et al.~\cite{wang2022efficient} Egeria solution by monitoring the rate by which the training loss changes, freezing layers when they reach the convergence criterion, and drop frozen layers from in both the back propagation phase and gradient exchange when data parallelism is used. It is important to note that Egeria periodically updates the reference model (on the CPU) to drive the layer freezing, yet does not actively try to remedy the load imbalance caused by layer freezing. The effect of load imbalance is particularly pronounced since earlier layers tend to be more frozen than later layers, i.e. the layer freezing is not uniformly occurring across the model. In comparison, \tool{} dynamically balances layer placement across GPUs, and in an orthogonal fashion, whenever the reference model used for freezing is updated.

\subsubsection{\textbf{Dynamic Sparse Flash Attention}}
We build on Pagliardini et al. dynamic sparse attention~\cite{pagliardini2023fast} in continual training by implementing a binding from PyTorch to the CUDA kernel provided by authors of the paper. The hash-based attention makes the causal attention mask irregular, i.e., we get blocked sparsity that is then leveraged by the Flash Attention. The irregular causal structures  caused by the hashing lead to different amount of blocks/tiles in different layers.

\subsubsection{\textbf{Early Exit}}
We adapt the early exit methods CALM~\cite{schuster2022confident} and ADP\-C~\cite{DBLP:conf/iclr/0003XWDS22} to observe the imbalance by peaking into the confidence measure prediction of CALM and ADP\-C. Since early exit happens at the later layers, we start our observation from the first layer at which tokens start to exit, and we assume all layers before than to have the same load. Early exit in particular benefits greatly from re-packing. That is since the change in control flow of the model happens in the later layers.

\subsubsection{\textbf{Mixture of Depths}}
We build on the MoDs work by Raposo et al.~\cite{Raposo2024MixtureofDepthsDA} by including in our GPT models we use in testing the small auxiliary MLP predictor that predicts whether that token will be among the top-k or not in the sequence. Similar to the case of MoEs, since the imbalance is dependent on the point the router takes the decision, i.e. in the forward pass, we rebalance in the back propagation phase where we attach the movement of layers to the pipeline parallelism scheme. Since the routing happens around the entire block, i.e., the routing applies to both forward MLPs and multi-head attention, we treat the skipped layers to be \textit{shadow} layers when redistributing the layers on workers.

\section{Evaluation}
\label{sec:evaluation}

\begin{figure*}[t]
\centering
\includegraphics[width=0.90\linewidth]
{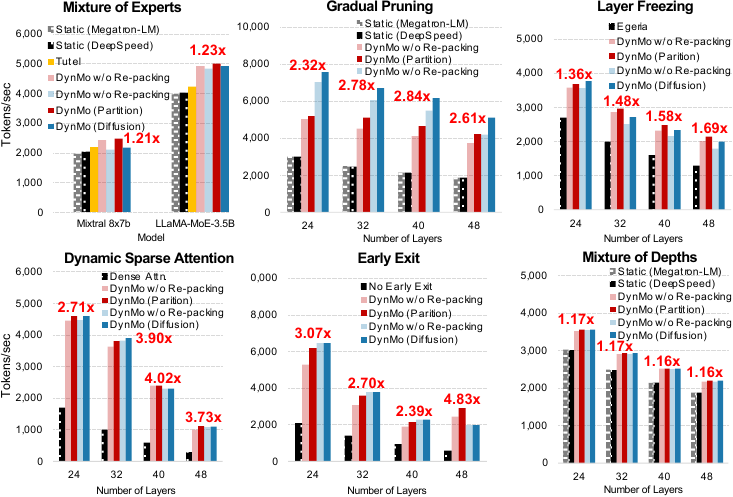}
\vspace{-0.2cm}
\caption{\rebuttal{Throughput of end-to-end training for six different example cases. \tool{} rebalances at regular intervals w/o prior knowledge of dynamism. MoE, Sparse Attn., and MoDs: invoke \tool{} every iteration. Pruning, layer freezing, and early exit: invoke \tool{} every 100s to 1,000s iterations. Speedup we report is the highest among balancing by number of parameters or layer execution time, divided by the highest among static Megatron-LM and DeepSpeed (or SoTA baseline, when available). MoEs and MoDs: we use 128 GPUs (16 nodes each with 4x H100s) in a hybrid of 8-way data parallel + 16-way pipeline). For gradual pruning, layer freezing, dynamic sparse attention, and early exit we use a total of 720 H100 GPUs (90 nodes each with 4x H100s) in a hybrid of 30-way data parallel + 24-way pipeline.}
}
\label{fig:end-to-end:multi}
\vspace{-0.2cm}
\end{figure*}

\label{sec:experiment_setup}

Experiments were primarily conducted on a supercomputer where each compute node consists of 2x AMD EPYC 9654 96-core/2.4GHz CPUs and 4x NVIDIA H100 SXM5 (80GB) GPUs. GPUs communicate with CPUs over PCIe Gen5 x16, and with each other via NVSwitch (NVLink34 x6). Nodes are interconnected via 4x 200Gbps InfiniBand NDR200 links. We used CUDA 12.6, OpenMPI 4.0.7, and PyTorch 2.3.1 with the NCCL 2.17.1 backend.

Models were trained on the Wikipedia dataset~\cite{wikidump}. For Mixtral 8x7B and LLaMA-MoE-3.5B, we perform continual training. All models use a sequence length of 2048, hidden size of 1024, and 32 attention heads. Unless otherwise specified, training runs for 10,000 iterations with micro-batch size 2 and batch size 64. In multi-node experiments, as we scale the number of GPUs, we proportionally increase the batch size to maintain four micro-batches per GPU, following the guidance in~\cite{huang2019gpipe} for optimal pipeline utilization.

We evaluated two dynamic load balancing algorithms, each with two configurations, across all six dynamic model scenarios. The first, Partition, uses DeepSpeed’s~\cite{rajbhandari2020zero} API and applies either parameter counts \textit{(Partition: by Param)} or decoder layer execution times \textit{(Partition: by Time)} to determine optimal layer placement using binary search and linear probing. The second, Diffusion, is a decentralized, iterative algorithm that balances load by minimizing variance across workers, with variants using either parameter counts {\em (Diffusion: by Param)} or execution times \textit{(Diffusion: by Time)}.

\begin{figure*}
\centering
\includegraphics[width=\linewidth]
{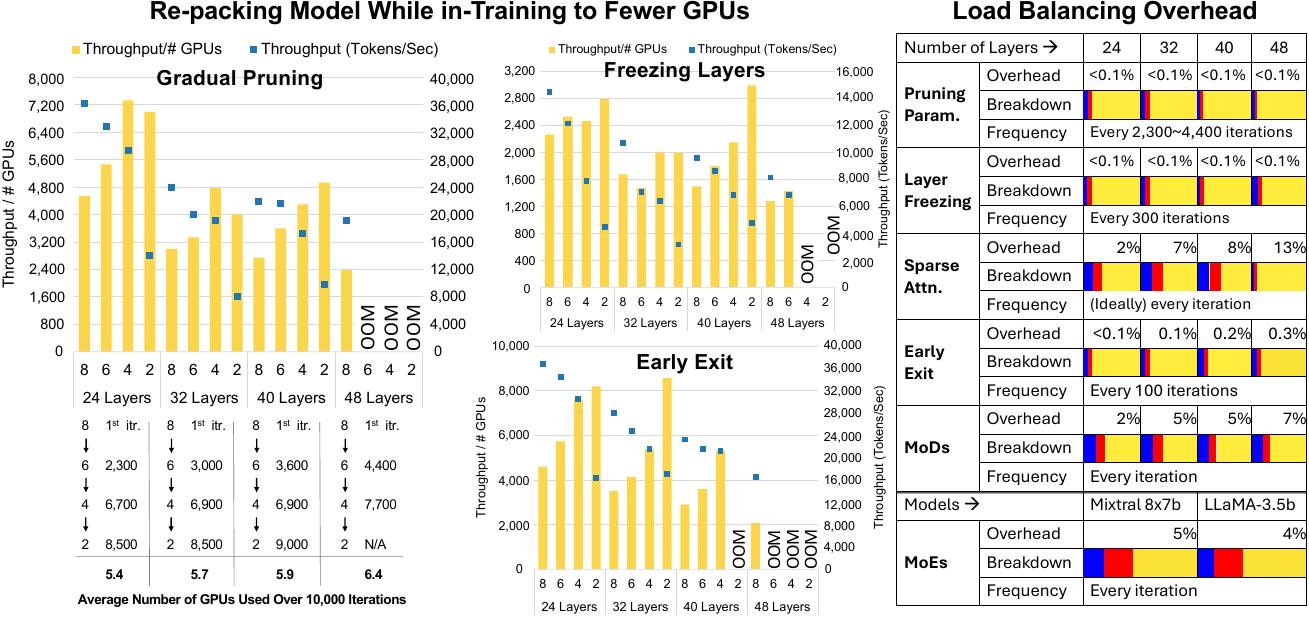}
\vspace{-0.2cm}
\caption{Re-packing the layers of GPT models into fewer GPUs as the model gets smaller due to dynamism. X-axis: Number of GPUs in pipeline parallelism 
using 90 nodes and up to 8 GPUs per node. 
Left Y-axis: throughput/number of GPUs shown with  yellow bars. Right Y-axis: throughput (tokens/sec) shown with blue squares. Below: we show the average number of GPUs needed throughout the gradual pruning training at which we dynamically re-pack (total 10,000 iterations). Right: load balancing overhead for example cases. The overhead reported is broken down to three things: profiling (in \colorbox{blue}{\phantom{XX}}), \tool{} load balancing algorithm (in \colorbox{red}{\phantom{XX}}), and  migration of layers between GPUs (in \colorbox{yellow}{\phantom{XX}}).}
\label{fig:repack-full}
\vspace{-0.2cm}
\end{figure*}
\subsection{End-to-end Training Throughput and Speedup}
All our throughput and speedup results include the load balancing overhead, unless specified otherwise. We trained GPT models~\cite{gpt-2} having different numbers of layers to determine the training throughput and speedup over static Megatron-LM and DeepSpeed (or SoTA baseline, when available). 

Figure~\ref{fig:end-to-end:multi} for MoEs, gradual pruning, and MoDs presents the highest throughput of two static and four dynamic load balancers. The first static balancer, Megatron-LM~\cite{shoeybi2019megatron}, evenly distributes layers across accelerators. The second static balancer, DeepSpeed~\cite{microsoft}, balances the number of parameters before training begins. In contrast, each of the two the dynamic load balancers (\emph{Partition} and \emph{Diffusion)} has two variants to redistribute the layers after each dynamism step: redistribute based on number of parameters or based on the layer execution time. Parameter-based balancers require profiling after the pruning step for memory usage information, while time-based balancers require profiling for memory usage and layer execution time information. We report the highest among both. 
Figure~\ref{fig:end-to-end:multi} for layer freezing, dynamic sparse attention, and early exit compare the two dynamic load balancers (\emph{Partition} and \emph{Diffusion)} over SoTA baseline that exists for those example cases. 

We observed that the use of layer execution time for dynamic load balancing, such as diffusion or partitioning, consistently outperforms parameter count-based implementations across all scales. In every scale, execution time-based dynamic balancers surpass the baseline static balancers. As seen in the figure, most of the speedup reported is attributed to balancing the load by redistributing the, and not due to the reduced communication when we re-pack: the speedup gain from re-packing is between 4$\sim$11\% of the entire speedup gain. In other words, even if we do not re-pack at, the speedup gain remains almost the same. Hence we treat re-packing as just a way to be efficient by using less GPUs, and not as a way to speed up imbalanced dynamic training. 

\textbf{Mixture of Experts.}
MoEs requires find-grained dynamism since the load vary from iteration to iteration. \tool{} shows more than 1.21x improvement on Mixtral 8x7b and LLaMA-MoE-3.5B in continual training. We do not change any of the hyperparameters from the original implementations. In addition to the Megatron-LM and DeepSpeed baselines, we also compare a highly MoE-tailored system: Tutel~\cite{tutel}. \tool{} significantly outperforms Tutel: 1.18x on Mixtral 8x7b and 1.21x on LLaMA-MoE-3.5B.

The improvement margins on MoEs and MoDs are in fact among the top in all six use cases, relative to how much for the bubble ratio was eliminated. \tool{} reduces the bubble ratios of MoEs and MoDs from 25\% to 8\% and 18\% to 4\%, respectively. As a result, the end-to-end training of two production models improve 1.21x on Mixtral 8x7b and 1.23x on LLaMA-MoE-3.5B. Those improvements would translate to significant cost savings, considering the huge cost of training those models (and other similar models).

\textbf{Gradual Pruning.}
The pruning region starts from iteration 3000 and continues until iteration 7000 and the model is pruned every 1000 iterations until the 90\% target sparsity is reached. This corresponds to sparsity levels of 52\%, 79\%, and 90\% after each pruning step. All other hyperparameters are the same as Megatron-LM. Using layer execution time for diffusion or partitioning dynamic load balancing outperforms the parameter count-based implementations in each scale, for up to 3.18x.

\textbf{Layer Freezing.}
\tool{} ourperforms the SoTA layer freezing tool Egeria~\cite{wang2022efficient}. We can observe two main points. First, the speedups of different load balancing algorithms over static algorithms are largely similar. This is mainly because the different algorithms tend to arrive at similar load balancing solutions when entire layers are frozen. Second, \tool{} shows increased speedup as the number of layers increases, particularly with diffusion, primarily because Egeria's overhead grows fast with the number of layers, while \tool{} overhead remains almost flat.

\textbf{Dynamic Sparse Attention.}
\tool{} achieves between 2.71x-4.02x speedup over the baseline dense attention (w/o sparsification). Dynamic sparse attention is the example case where \tool{} is most efficient at removing the pipeline bubbles. That is since using layer time execution, which fluctuates a lot in dynamic sparse attention, enables effective redistribution of the layers.

\textbf{Early Exit.}
\tool{} achieves more than 4x on average over the baseline w/o exit, i.e. when all tokens pass through the entire model. Similar to dynamic sparse attention, early exit benefits the most from \tool{} due to the big variance in load between earlier and later layers. 

\textbf{Mixture of Depths.}
Like MoEs, MoDs layer loads vary from iteration to iteration. \tool{} shows 1.17x improvement on the baseline in continual training. We suspect MoDs will in the future be able to benefit more from custom load balancers that leverage the knowledge of how MoEs is used in hybrid with MoDs.

\subsection{Overhead of Load Balancing}
Figure~\ref{fig:repack-full} (right) reports the load balancing overheads. This includes both the load balancing decision and the actual transfer of the parameters and other data of the layers to be sent or received, e.g., row offsets and column indices in CSR format for gradual pruning, and gradients in the case of MoEs and MoDs. The overhead is generally negligible, hence giving the opportunity for the use of \tool{} in other forms of dynamic models, and not just the six example cases in this paper.

\subsection{Re-packing Models to Fewer GPUs}
In the re-packing experiments, the training starts with 8 GPUs per node in pipeline parallelism. 
After a dynamism step, \tool{} attempts to re-pack the total workload into fewer GPUs while satisfying the memory capacity constraints. Figure~\ref{fig:repack-full} reports the throughput/number of GPUs for each model size where the model is packed into 6, 4, and 2 GPUs. The 8 GPU setting for each model size serves as a baseline where there is no re-packing. This measurement also corresponds to the performance per dollar metric as the cost is directly proportional to the number of GPUs used in training. 

We observe that in all model scales, re-packing can allow the training to be continued with fewer GPUs which may result in significant cost savings. For example, in gradual pruning we reduce the GPU count from 8 to an average of 5.8 GPUs while sustaining the training throughout.









\section{Related Work}
\subsection{Load Balancing Model-Parallel DNNs}
\textbf{Layer-wise load balancing.}
Layer-wise balancing techniques operate at the granularity of layers rather than individual operators. DeepSpeed~\cite{microsoft} provides three partitioning methods to balance stage workloads.  The parameters method (i) aims to equalize the number of parameters across stages,  while the uniform method (ii) distributes layers evenly.  The regex method (iii) selectively distributes layers that match a given pattern, such as transformer layers. Similarly, He et al.\cite{he2021pipetransformer} balance stages based on parameter counts, and Narayanan et al.\cite{narayanan2021efficient} assign an equal number of transformer layers to each stage. However, none of these approaches use actual layer execution time to guide partitioning. \tool{} leverages DeepSpeed’s partitioning scheme by supporting both parameter counts and layer execution times for load balancing, and also includes a diffusion-based algorithm as a built-in option.

{\textbf{Load balancing via graph partitioning.}}
Graph partitioning-based load balancing schemes identify atomic operations in the model and represent them as nodes in a directed acyclic graph (DAG), where edges capture data dependencies. Tanaka et al.\cite{tanaka2021automatic} partition the DAG in three phases: identifying atomic operations, grouping them into blocks based on computation time, and combining these blocks into final partitions using DP. Qararyah et al.\cite{qararyah2021computational} form disjoint clusters by identifying critical paths and mapping them to devices using an algorithm that accounts for both communication minimization and temporal load balancing. Both approaches rely on profiling before training and partition only once.

{\textbf{Load balancing in MoE Models.}}
MoE\cite{jacobs1991adaptive} models contain many sub-networks (experts) where a router allocates inputs to top-k experts. At scale, experts are distributed across devices. Lepikhin et al.~\cite{lepikhin2020gshard} defines an expert's capacity to limit the maximum number of tokens that can be processed by an expert to achieve workload balance. Fedus et al.~\cite{fedus2021switch} route each token to only one expert and use the same expert capacity for restrictions. Lewis et al.~\cite{lewis2021base} employ an auction algorithm~\cite{Bertsekas1992} to solve the token-to-expert assignment problem. This line of work is different from ours as their aim is to balance workload in the feed-forward network while our work aims to balance all layers of the transformer model.

\subsection{Re-Packing}
PipeTransformer~\cite{he2021pipetransformer} offers an elastic pipelining system for freeze training where some of the layers of the model are frozen during the training. PipeTransformer packs the remaining active layers into fewer GPUs and creates pipeline replicas if possible. When PipeTransformer receives a notification for layer freezing, it attempts to divide the number of GPUs by 2 subject to the memory capacity constraints. On the other hand, our work \tool{} can pack to an arbitrary number of GPUs specified by the user. 
PipeTransformer uses the parameter size as a proxy to estimate the memory usage while \tool{} uses the actual memory usage from the profiling step before load balancing. Finally, PipeTransformer is only capable of re-packing, and not load balancing. \tool{} can do both. 
\section{Conclusion}
\label{sec:conclusion}

Research on dynamic models will not deliver practical impact unless there is a platform from which those models can be made efficient. 
\tool{} serves as that platform, offering a load-balancing system and a re-packing solution specifically designed for dynamic models whose worker loads vary during training. This leads to improved efficiency and faster end-to-end training times. Empirical results on LLMs, demonstrated across six representative dynamic model cases, show that \tool{} significantly improves training throughput compared to existing approaches. Moreover, it enables elastic GPU utilization, resulting in additional cost savings. As dynamic models become increasingly prevalent, dynamic load distribution will be critical for scalable and efficient training.
 

\begin{acks}
Authors from Koç University has received funding from
the European Research Council (ERC) under the European Union’s Horizon 2020 research and innovation programme (grant agreement No 949587).
\end{acks}


\bibliographystyle{ACM-Reference-Format}
\bibliography{sample-base}

\end{document}